\documentclass[12pt]{article}
\usepackage[latin1]{inputenc}
\usepackage{makeidx,color}
\usepackage{amsfonts}
\usepackage{graphicx}
\usepackage{amssymb,amsmath,cite}
\usepackage{subfigure}
\usepackage[margin=9pt,font=small,labelfont=bf,textfont=it,format=hang,labelformat=simple]{caption}
\date{}
\begin{document}

\title{Photon statistics of radiation in an incoherently pumped three-level cascade system}
\author{Shaik Ahmed, Preethi N. Wasnik,\\ Suneel Singh and P. Anantha Lakshmi*\\\\
School of Physics, University of Hyderabad\\ Hyderabad - 500046, India\\\\
*palsp@uohyd.ernet.in}
\maketitle

\begin{abstract}

We study the intensity-intensity correlations  of the radiation emitted on probe transition in a three level cascade electromagnetically induced transparency (EIT) scheme.  By applying an incoherent pump, we also monitor further changes in the characteristics of the emitted radiation.   It is found  that application of even a very weak incoherent pump can significantly alter the characteristics of the emitted radiation, even though the EIT characteristics remain unaltered.   Our study demonstrates that for certain range of parameter values, the two photon correlation function in the probe transition oscillates between classical and non-classical domains.
\end{abstract}

PACS number(s) 42.50.Ar, 42.50.Gy, 42.50.Hz

\section{INTRODUCTION}

 It is well known that the first order correlation function gives information about the absorption and emission spectra of the radiation emitted from an atomic system under the influence of external radiation fields.  On the other hand,   a study of second order (intensity - intensity) correlation function of the radiation field determines its photon statistics\cite{r0}, which enables one to gain  insight into the nonclassical nature of the radiation field.
 Study  of intensity correlations of emitted radiation from multilevel systems driven by external fields, in different environments, is of  tremendous interest in recent times.  Several studies of two photon correlations in driven three level systems in various ($\Lambda$-, V- and cascade) configurations exist.
Huang et al \cite{r1} have studied the intensity autocorrelations in  fluorescence from the upper and lower transitions of  a strongly driven three level cascade system.   They  found  additional resonances in the upper transition which was attributed to the influence of spontaneous decay between the lower transition levels.
In a different study \cite{r2,r3}, Ficek et al.  investigated the two photon correlations in a three level V-type system, with perpendicular dipole moments.  A DC field was applied to create an additional two photon coherence between the upper two  levels.
Two photon correlations have also been employed as a tool to probe the vacuum induced coherences by  \cite{r4} Das et al.   They have demonstrated interference phenomena in the second-order correlations of the fluorescence light from ion pairs.  From their work it is inferred that higher order photon  correlations could be present  even in situations where first-order interferences do not appear \cite{r5}.
In another study,  the effect of vacuum-induced coherence on the photon correlation\cite{r10} in an equi-spaced three level cascade system  has been reported.
Several of these studies aim at investigating the modification in the characteristics of the emitted light,  because a study of the first and higher order correlation functions of the emitted radiation fields gives information about the nature of the light.  In addition, such a study can further be exploited to probe other quantum features.

Thus, two photon correlation spectroscopy is found to be a very  effective tool in determining the nature of the radiation fields  emitted from the  atomic system under different excitation mechanisms.  The   characteristics of the emitted light vary,  depending on the nature of excitation, various incoherent processes that influence  the system and any other external perturbers.
On the other hand, electromagnetically induced transparency (EIT)\cite{r6}  and lasing without inversion (LWI) \cite{r7}are  well studied phenomena  and still are of current  research interest.
It would be interesting  from the experiment point of view to look at the nature of emitted radiation in the EIT scheme and also in the LWI  configuration, the latter of which involves application of an incoherent pump.  The present work deals with such a study.

For this purpose, we have considered  the atomic system in a three level cascade  configuration, which can be related to a typical  EIT scheme in any of the alkali atoms\cite{r6}.  The lower transition of this cascade system is driven by a weak probe field while  the upper transition is driven by a strong pump field.
 We do not  impose any restrictions on the energy level separations of the upper and lower transitions.  Our model incorporates the effect of  incoherent pumping from the ground level  to both the excited levels,  in addition to the spontaneous emission decays.

The radiation field emitted, in the far field zone,  in both the pump and probe transitions can be  expressed in terms of the atomic operators.   Using the quantum regression theorem, the multi-time correlation functions can be written in terms of the expectation values of the atomic operators in steady state multiplied by the function of the delayed time, which comes from the assumption of statistical stationarity of the fields.

In particular,  our studies  demonstrate that application of a very weak incoherent pump can significantly alter the characteristics of the emitted radiation.  For suitable values of the parameters, the two photon correlation function in the probe transition oscillates between non classical and classical behaviour.

 The organisation of this paper is as follows.  In section II the mathematical formulation for obtaining the equations of motion for the atomic density matrix  elements is given.  This is followed by a brief description of the intensity-intensity correlation functions of the fluorescent fields  in terms of the correlation functions of the atomic operators.   In section III, we present a detailed analysis of the behaviour of the above system in terms of the absorption spectra and the corresponding intensity-intensity correlation function, in particular, for the probe transition.

\section{FORMULATION}
\begin{figure}[ht]
\centering
\centering
\includegraphics[height=5 cm,width=5 cm]{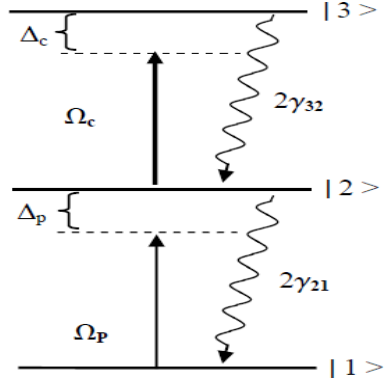}
\caption{ Three level cascade system - energy levels  and interaction scheme}
\label{fig:One}
\end{figure}

 We model the atom as a three level system in cascade configuration.  The atomic level scheme is shown in Figure \ref{fig:One}.  The top level $|3\rangle $ (energy $E_3$) decays spontaneously with a rate $ 2\gamma_{32} $  to the  intermediate level $|2\rangle $ ( energy $ E_2 $) which decays spontaneously with a rate  $ 2\gamma_{21} $ to the   ground  level $|1\rangle $  (energy $ E_1 $ ).
  The pump and probe laser fields drive the transitions $|2\rangle $-$|3\rangle $ and $|1\rangle $-$|2\rangle $ with Rabi frequencies $\Omega_c$ and $\Omega_p$ respectively.  In addition, the upper levels $|3\rangle $ and $|2\rangle $ may be incoherently pumped from the ground level $|1\rangle $, at the rate $\Lambda_{13}$ and $\Lambda_{12}$ respectively.

The Hamiltonian for this system in the electric dipole approximation is
\begin{eqnarray} H = H_0 & - & \left(  |3\rangle \langle 2|  \vec\mu_{32}  +  |2\rangle \langle 3|  \vec\mu_{23} \right) \cdot \left(\vec E_c e^{-i \omega_c t }+ c.c. \right) \cr \cr
&  - &  \left(  |2\rangle \langle 1|  \vec\mu_{21} +   |1\rangle \langle 2|  \vec\mu_{12}  \right) \cdot \left( \vec E_pe^{-i \omega_p t }+ c.c. \right)
\end{eqnarray}

The equation of motion for the density operator $\rho$  of the atomic system is

\begin{equation}
i\hslash \frac{\partial\rho}{\partial t}=-[H,\rho]+\mathcal{L} \rho
\end{equation}
where the second term accounts for spontaneous emission from the upper levels. The equation of motion of the density matrix elements of the system after making the rotating wave approximation are obtained as

{\begin{align}
&\frac{d\rho_{11}}{dt}= -2(\Lambda_{12}+\Lambda_{13})\rho_{11}+2\gamma_{21} \rho_{22}+i\Omega_p(\rho_{21}-\rho_{12})\cr
&\frac{d\rho_{22}}{dt}= 2\Lambda_{12}\rho_{11}+2\gamma_{32} \rho_{33} - 2\gamma_{21} \rho_{22} - i\Omega_p(\rho_{21}-\rho_{12}) - i\Omega_c(\rho_{23}-\rho_{32})\cr
&\frac{d\rho_{33}}{dt}= 2\Lambda_{13}\rho_{11}-2\gamma_{32} \rho_{33}+i\Omega_c(\rho_{23}-\rho_{32})\cr
&\frac{d\rho_{12}}{dt}=-( \gamma_{21}+\Lambda_{12}+\Lambda_{13}+i \Delta_{p})\rho_{12}+i\Omega_p(\rho_{22}-\rho_{11})-i\Omega_c \rho_{13}\cr
&\frac{d\rho_{23}}{dt}=-( \gamma_{21}+ \gamma_{32}+i \Delta_{c})\rho_{23}+i\Omega_c(\rho_{32}-\rho_{22})+i\Omega_p\rho_{13}\cr
&\frac{d\rho_{13}}{dt}=-[ \gamma_{32}+(\Lambda_{12}+\Lambda_{13})+i (\Delta_{p}+  \Delta_{c})]\rho_{13}+i\Omega_p\rho_{23}-i\Omega_c \rho_{12}
\end{align}}
together with  $\rho_{ij} = \rho_{ji}^{*}$. In these equations,  $\Delta_{p}=\omega_{21}-\omega_p $ and $\Delta_{c}=\omega_{32}-\omega_c $ are the frequency  detunings of probe and pump laser fields respectively.  The Rabi frequencies for the probe  and pump  transitions respectively are $\Omega_p= \frac{\vec\mu_{21} \cdot \vec E_p}{\hbar}$ and $\Omega_c= \frac{\vec\mu_{32} \cdot \vec E_c}{\hbar}$ (which are assumed to be real with no loss of generality).

The normalised two-time intensity correlation function, expressed in terms of the atomic populations in steady state and the evolution operator of the density matrix \cite{r1,r3}  is given by
\begin{equation}
G_{ij}(\tau) = \frac{\langle : I_i(t) I_i(t+\tau):\rangle}{\langle|i\rangle \langle i|\rangle \langle| j\rangle \langle j|\rangle} = \frac{P_{i+1\rightarrow j}(\tau)}{P_j} .
\end{equation}
Here $P_j$ is the steady state population in $|j\rangle$, and $P_{i+1\rightarrow j}(\tau)$ is the probability of population transfer, within the time interval $\tau$,  from  the final state of the first emission $|i+1\rangle$, to the initial state of the second emission $| j\rangle$.  These quantities can be evaluated\cite{r1,r3} using the set of equations (3) for the density matrix elements.
Our interest here is to study the intensity - intensity  correlation of the emitted radiation field in the probe transition, $G_{22}(\tau)$, as a function of different parameters.  
  In the next section, we  present numerical results for certain representative set of parameters, for the absorption profile as well as the corresponding  intensity - intensity correlation function of the radiation field emitted in the probe transition.  
 \section{ RESULTS AND DISCUSSION}

In this section,we present and discuss the  results for probe field emission characteristics, namely  the intensity - intensity correlations under the conditions of electromagnetically induced transparency (EIT) \cite{r6} and possibly lasing without inversion (LWI) \cite{r7},   through incoherent pumping of the upper level($|2\rangle$)  of the probe transition.

For this purpose,  the intensity correlation function $G_{22}(\tau)$ corresponding  to the emission in the probe transition $|2\rangle  \rightarrow |1\rangle$ along with its absorption profile are evaluated numerically as a function of various field and atomic parameters, viz.  the pump field Rabi strength ($\Omega_c$), the rate of incoherent pumping ( $\Lambda_{12}$ ) into the upper level of the probe transition and the  spontaneous decays in the probe ($\gamma_{21}$ )  and pump ($ \gamma_{32}$) transitions.

The parameters chosen here correspond to those used in the  experimental studies of EIT in three - level cascade system, for example  the $3S_{1/2} \rightarrow 3P_{1/2} \rightarrow 4D_{3/2}$ transition in a sodium atom.  For this transition, the level separation wavelengths are $\lambda_{21} = 589.6$ nm, $\lambda_{32} = 568.3$ nm.  The corresponding spontaneous emission decay rates are  $2\gamma_{21} = 2 \pi$(10  MHz) and $2\gamma_{32} = 2 \pi$(1.6 MHz) \cite{r8}.

\begin{figure}[ht]
\centering
\subfigure[]{
\includegraphics[width=6.5 cm,height=8 cm]{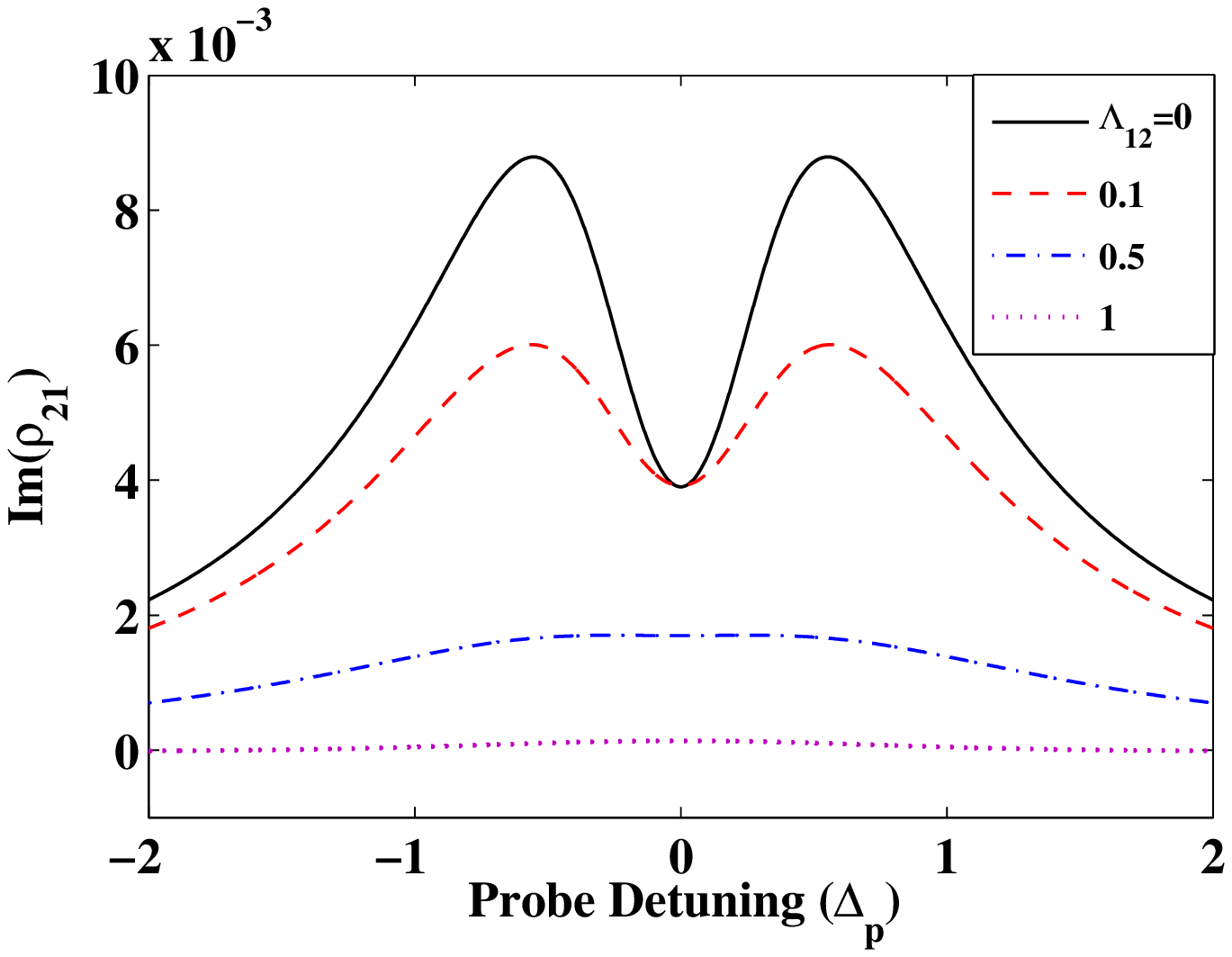}
\label{fig:Two a}
}
\subfigure[]{
\includegraphics[width=6.5 cm,height=8 cm]{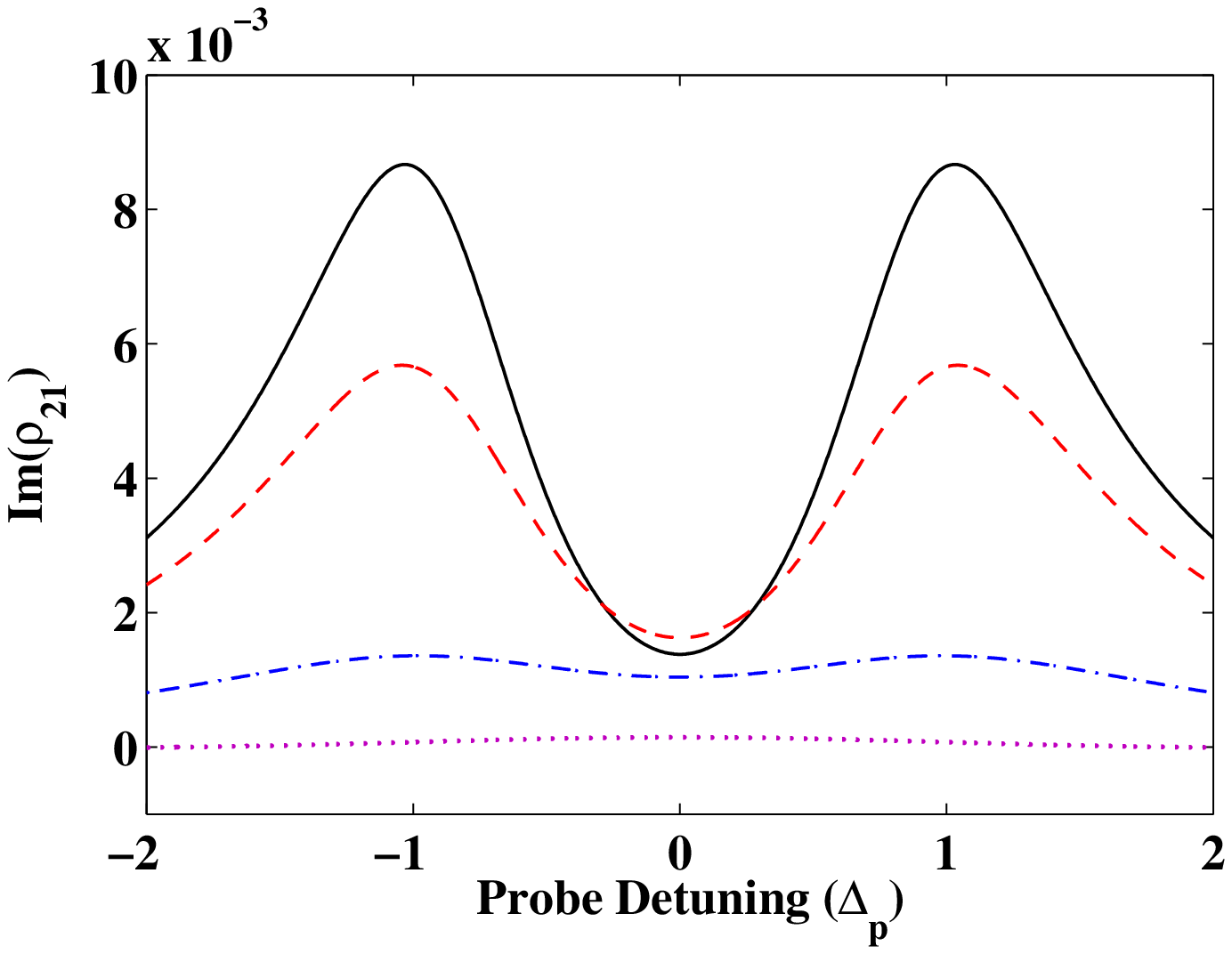}
\label{fig:Two b}
}
\caption[]{\textit{Imaginary  part of $ \rho_{21} $}  (a)  $ \Omega_{c}=0.5 $ and  (b) $ \Omega_{c}= 1$  for unequal spontaneous emission rates $ \gamma_{32}=0.16 $. The values of incoherent pumping  $\Lambda_{12}$ for each curve are shown in the legend.  The probe field Rabi frequency $ \Omega_{p}=0.01$ and  the incoherent pumping rate $ \Lambda_{13}=0$.  All parameters ($\Delta_p, \Omega_c, \Omega_p, \gamma_{32}, \Lambda_{12}$ and  $ \Lambda_{13}$)
used here and in subsequent figures  are in units of $\gamma_{21}$.   }  
\label{fig:Two}
\end{figure}
\begin{figure}[ht]
\centering
\subfigure[]{
\includegraphics[width=6.5 cm,height=8 cm]{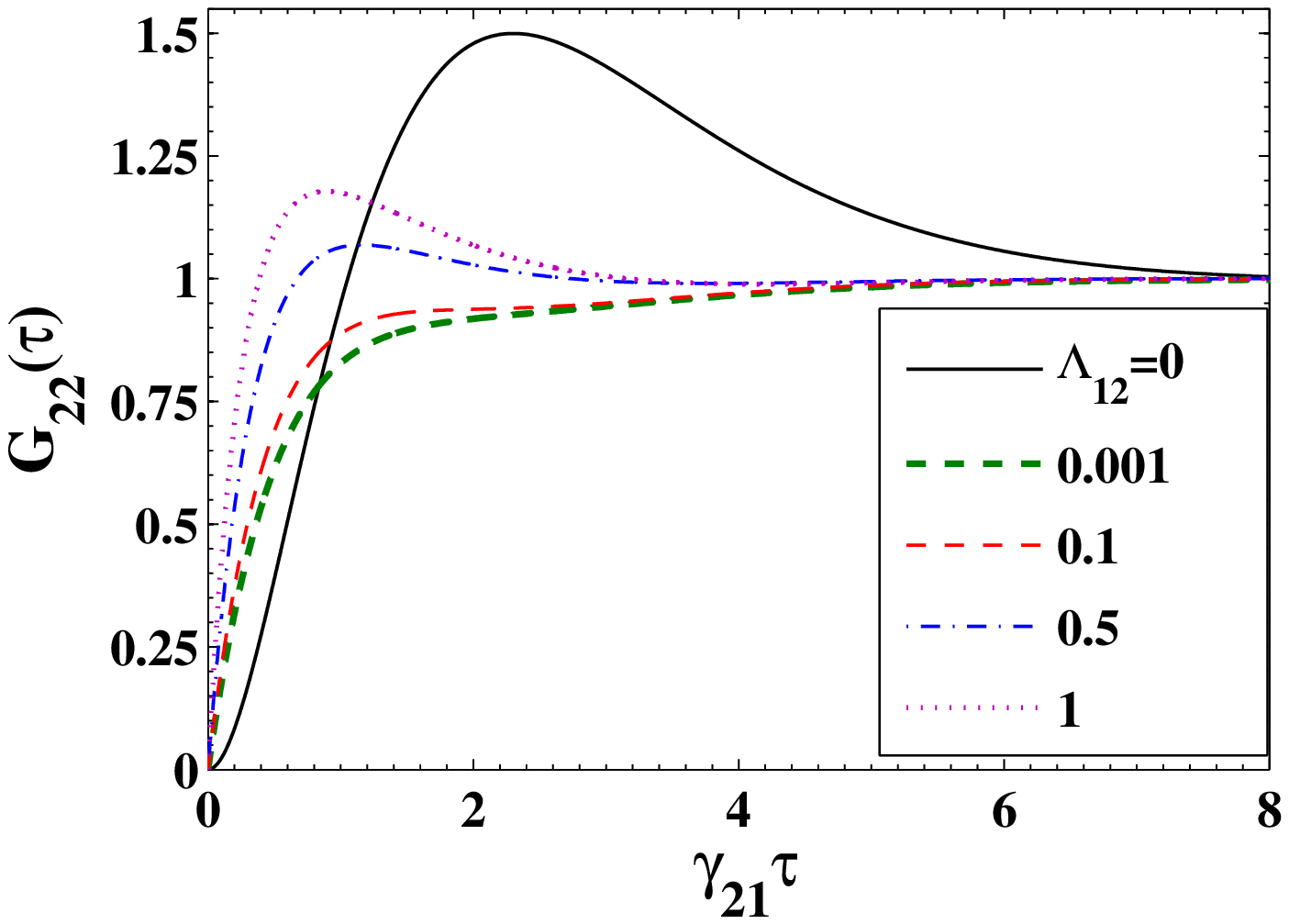}
\label{fig:Three a}
}
\subfigure[]{
\includegraphics[width=6.5 cm,height=8 cm]{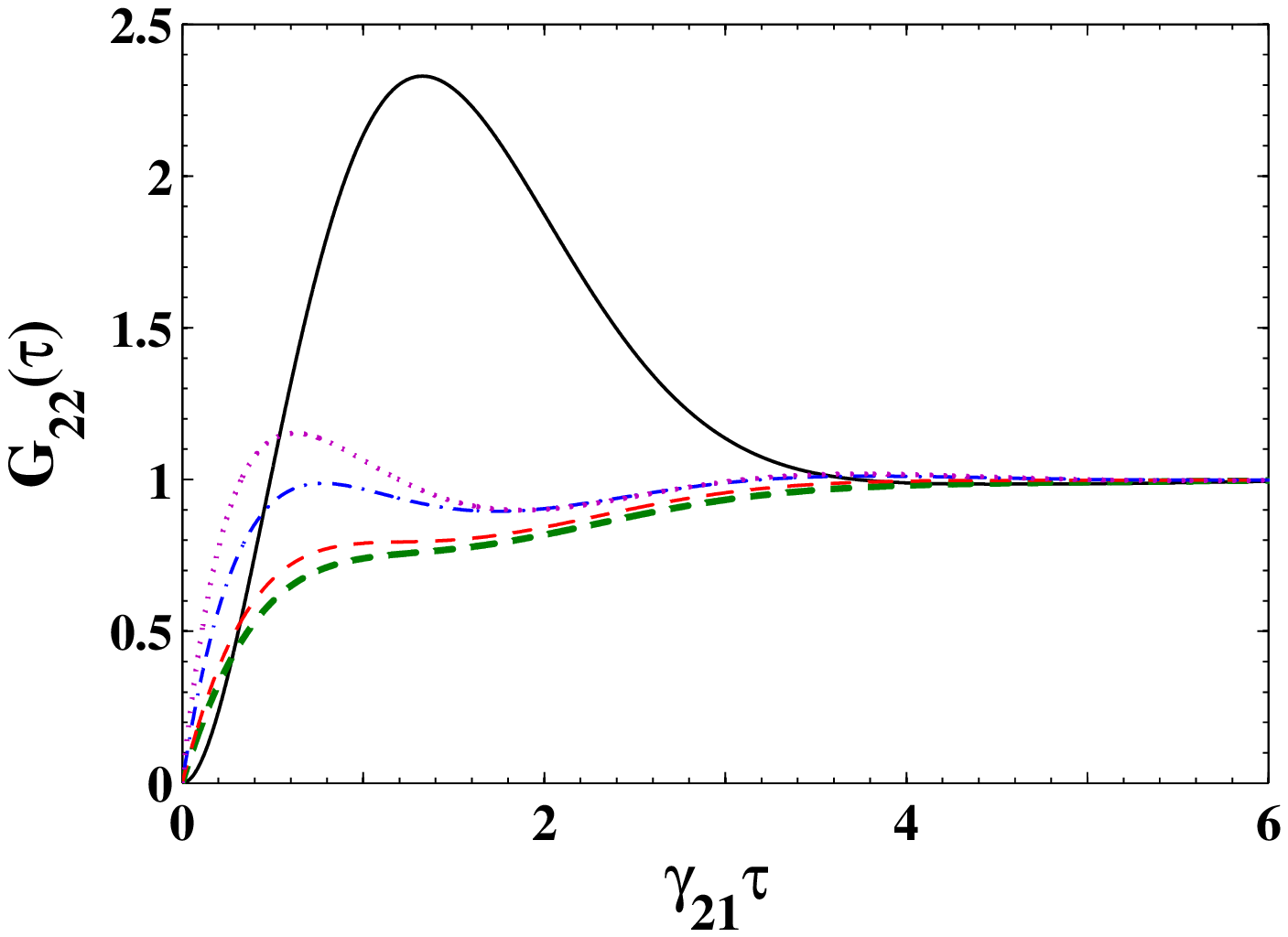}
\label{fig:Three b}
}
\caption[]{\textit{The probe field intensity correlation function $G_{22}(\tau)$  (a)  $ \Omega_{c}=0.5 $ and  (b) $ \Omega_{c}= 1$ for unequal spontaneous emission rates $\gamma_{32} = 0.16$ and for  the values of $\Lambda_{12}$ as shown in the legend.   Other parameters are same as in Figure 2.} }
\label{fig:Three}
\end{figure}

\begin{figure}[ht]
\centering
\subfigure[]{
\includegraphics[width=6.5 cm,height=8 cm]{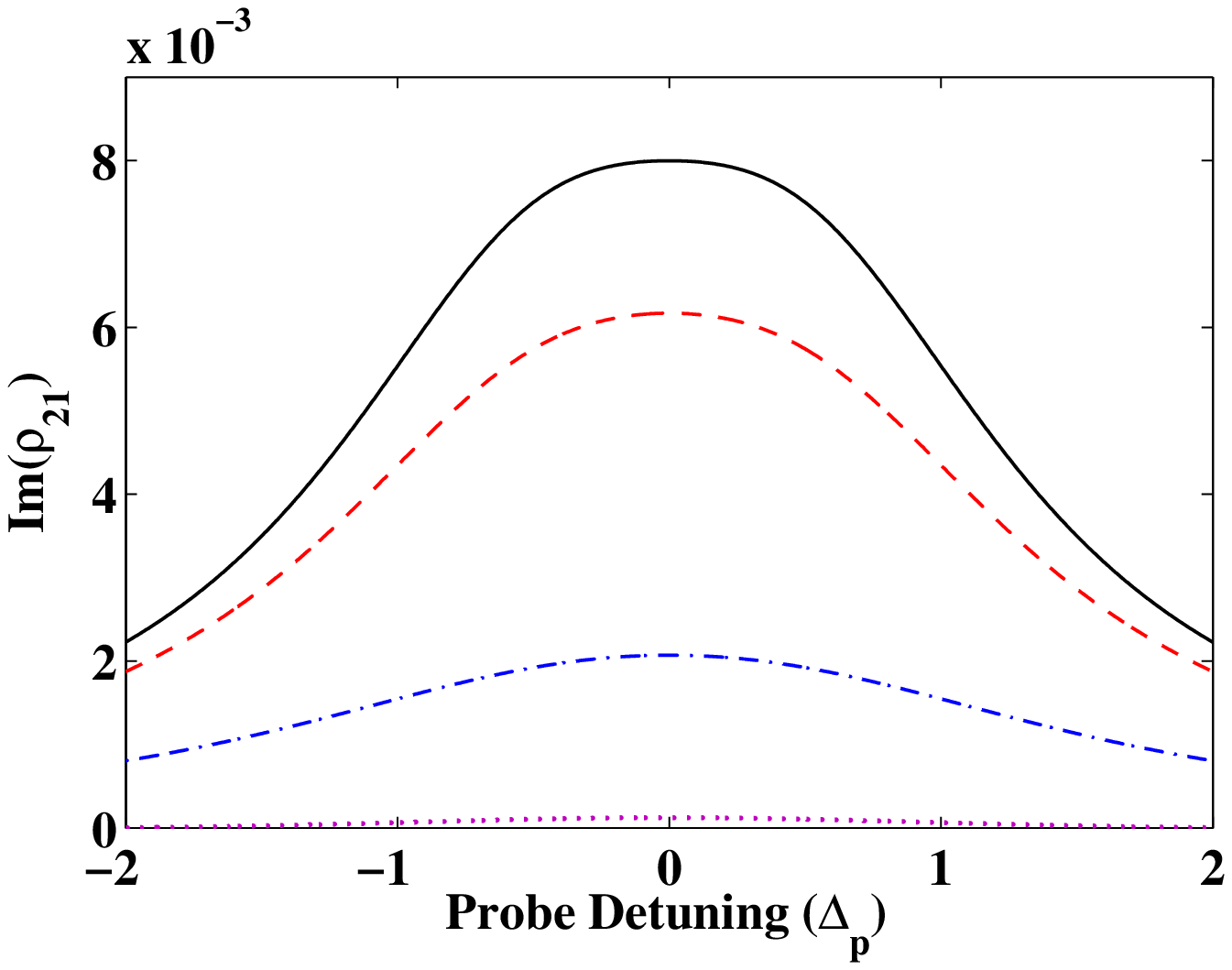}
\label{fig:Four a}
}
\subfigure[]{
\includegraphics[width=6.5 cm,height=8 cm]{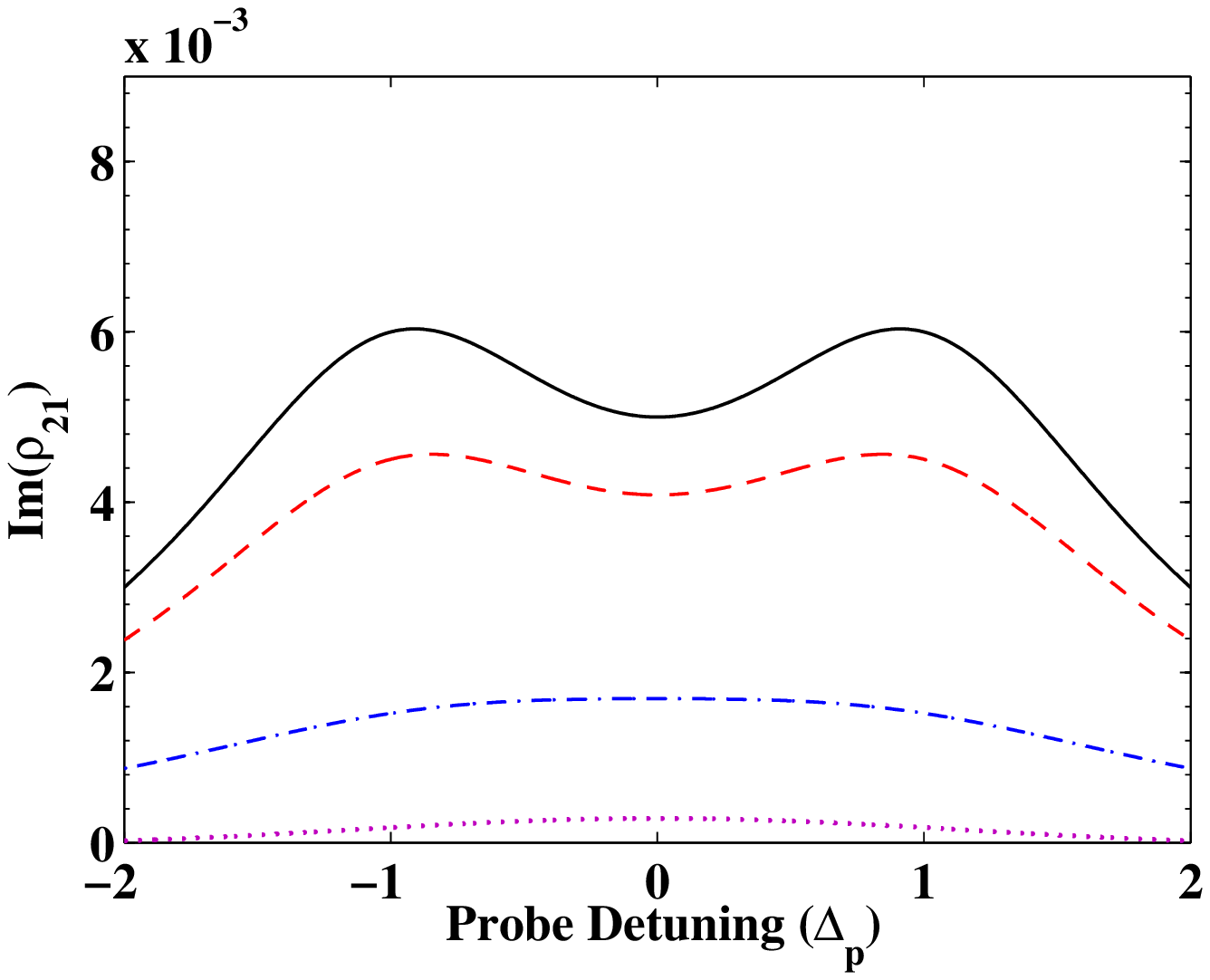}
\label{fig:Four b}
}
\caption[]{\textit{Imaginary  part of $ \rho_{21} $  (a)  $ \Omega_{c}=0.5 $ and  (b) $ \Omega_{c}= 1$  for equal spontaneous emission rates $\gamma_{32}=\gamma_{21}$. Other parameters are same as in Figure 2.} }
\label{fig:Four}
\end{figure}

In the EIT regime,    the control field Rabi frequency $\Omega_c$ is strong compared to the probe Rabi frequency $\Omega_p$.  Hence, we have  fixed the probe field Rabi frequency   at $\Omega_p = 0.01\gamma_{21}$ and used two typical  control Rabi field strengths viz.,  $\Omega_c$ = $\gamma_{21}/2$ and  $\gamma_{21}$ respectively in figures \ref{fig:Two a} and \ref{fig:Two b} to demonstrate the EIT characteristics of the probe field as  a function of the probe detuning $\Delta_{p}$.  In addition, we also explore in these figures, the effect of  incoherent pumping   into the upper level of the probe transition ($|2\rangle$)  through the ground level  ($|1\rangle$), denoted by the parameter $\Lambda_{12}$.

It is seen from figures \ref{fig:Two a} and \ref{fig:Two b} that for smaller incoherent pump rates  $\Lambda_{12} << \gamma_{21}$, there is no appreciable change in the EIT dip at the line centre and it is nearly identical to the case  when there is no incoherent pumping.  However  the dip vanishes  as the incoherent pumping rate increases  but the overall probe absorption reduces  further over a wider frequency range.   In order to understand the absorption reduction  with increasing incoherent pumping rate  $\Lambda_{12}$, we have also examined the steady state population difference ($\rho_{22} - \rho_{11}$) between the upper and lower levels of the probe transition (figure not shown here).   It is found that, for  values of  $\Lambda_{12}  \ge  0.5 \gamma_{21}$,  the  vanishing of EIT dip  and absorption reduction occurs due to saturation of the populations of the levels involved in the probe transition.  We find no evidence of LWI in the parameter regime studied here.

 \begin{figure}[ht]
\centering
\subfigure[]{
\includegraphics[width=6.5 cm,height=8 cm]{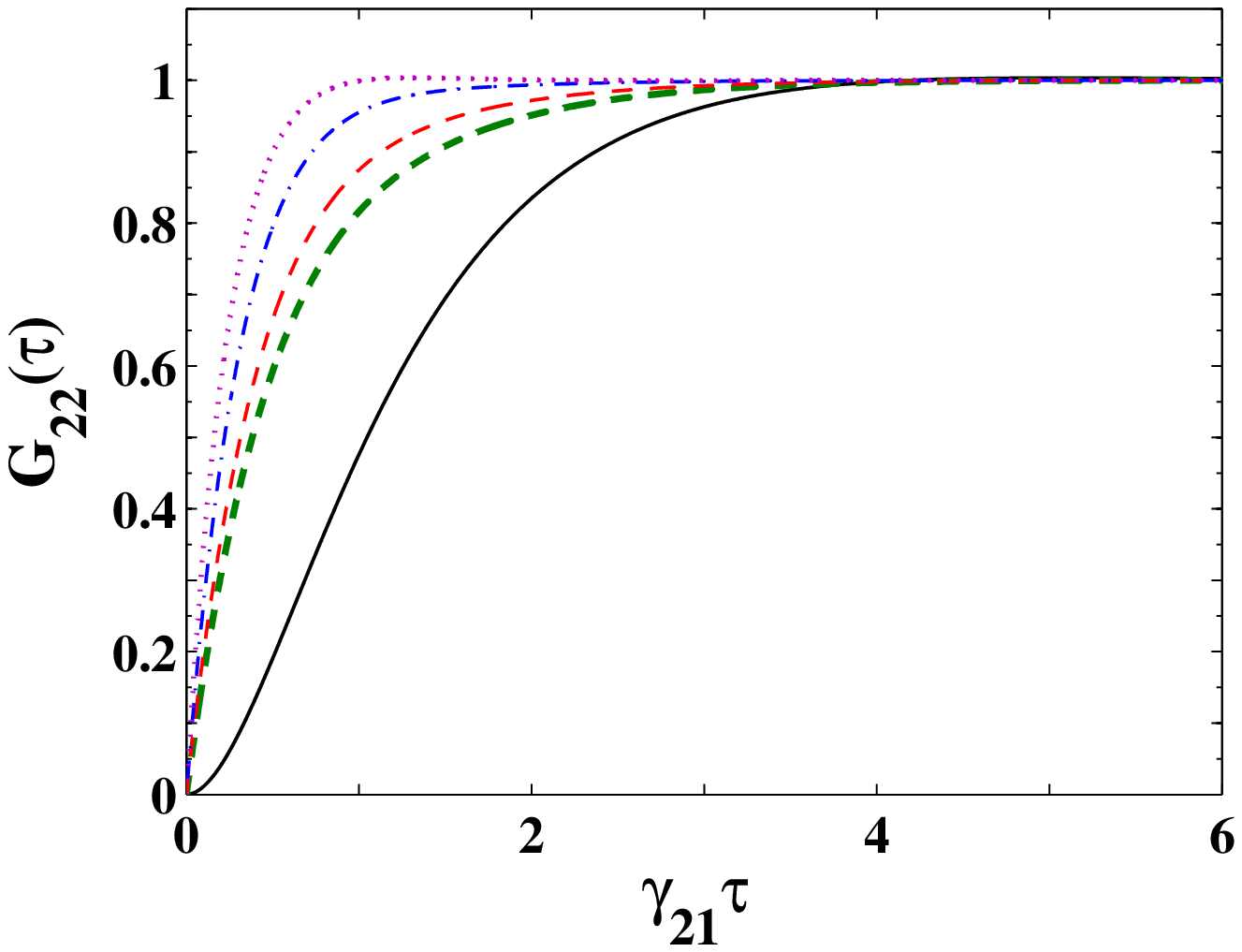}
\label{fig:Five a}
}
\subfigure[]{
\includegraphics[width=6.5 cm,height=8 cm]{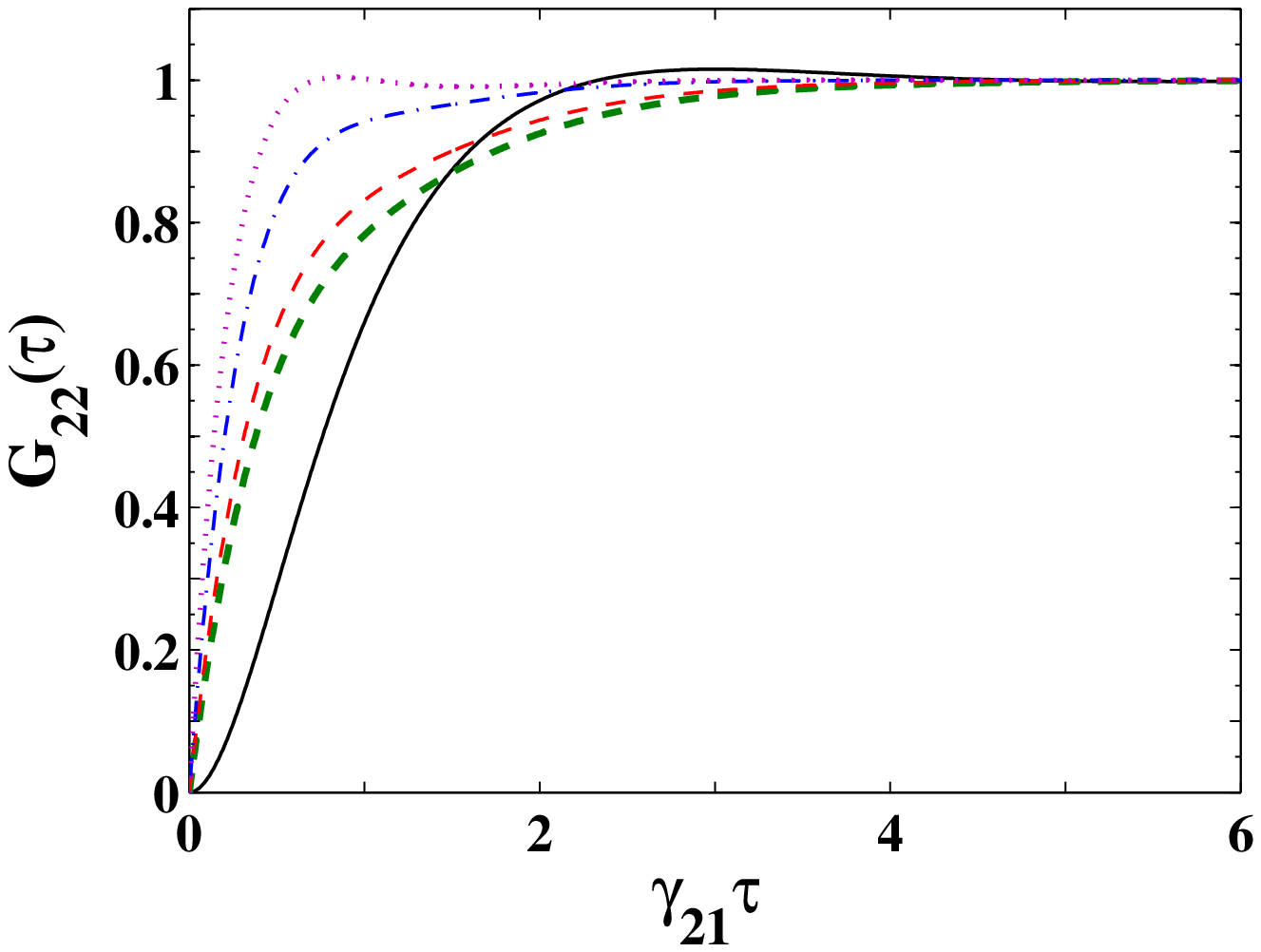}
\label{fig:Five b}
}
\caption[]{\textit{ The probe field intensity correlation function $G_{22}(\tau)$  (a)  $ \Omega_{c}=0.5 $ and  (b) $ \Omega_{c}= 1$  for equal spontaneous emission rates $\gamma_{32}=\gamma_{21}$.} Other parameters are same as in Figure 3.}
\label{fig:Five}
\end{figure}

The intensity correlation  function $G_{22}(\tau)$  of the probe transition, corresponding to the same parameters as in figures \ref{fig:Two a} and \ref{fig:Two b} is illustrated in figures \ref{fig:Three a} and \ref{fig:Three b} respectively.
These figures reveal that  incoherent pumping $\Lambda_{12}$  has a significant influence on the emission characteristics of the probe field.  In the absence of the incoherent pumping ($\Lambda_{12}=0$),  we find that  within a correlation time $\tau  \sim 1/ \gamma_{21}$, the correlation function rapidly crosses the unity mark  and enters the classical domain  wherein $G_{22} (\tau) > 1$ and remains classical for all later times.  On the other hand, even a very weak incoherent pumping causes  the correlation function to exhibit non classical behaviour.
This can  clearly be seen from the curves for  non zero values of $\Lambda_{12}$ ranging from $0.001\gamma_{21}$ to $0.1\gamma_{21}$.   Note also that,  for the same values of  $\Lambda_{12}$,  there is hardly any variation in the EIT profile  as can be  seen from figures \ref{fig:Two a} and \ref{fig:Two b}.   At higher values of $\Lambda_{12}  \ge  \Omega_c$, the correlation function re-enters the classical regime.

 Comparison between the probe intensity correlation  function, for two different values of the control field strength, plotted in figures \ref{fig:Three a} and \ref{fig:Three b} shows that for the larger of the two values of the control field, namely $\Omega_{c} = \gamma_{21}$, the functions exhibits stronger classical behaviour before settling into the steady state of uncorrelated random light ($G_{22} (\infty) =1$).
In both cases,  it is observed that the re-entry into the  nonclassical regime (for  a range of  values of the time delay $\tau$) occurs whenever $\Lambda_{12}\ge \Omega_c$.  The correlation function also is seen to exhibit oscillations  for this particular value of the control field. This oscillatory behaviour can be understood as resulting from the Rabi oscillations of the populations of the levels participating in the transition.  These Rabi oscillations essentially populate and depopulate the upper level of the probe transition periodically giving rise to oscillatory behaviour in the correlation function.

       To further clarify the influence of the spontaneous decay rates in the system on the intensity - intensity correlation characteristics,  we have also considered equal decay rates  ($\gamma_{21}  =  \gamma_{32}$),   as it may be relevant for  many  atomic systems.

       A comparison of the EIT feature in both cases reveals  the following.  For unequal spontaneous decay rates,  we have already seen from figures \ref{fig:Two a} and \ref{fig:Two b} that the EIT feature is more pronounced with  absorption reduction of around  90\% at the line center.   On the other hand, we find that for the case of equal spontaneous decays [cf.  figures \ref{fig:Four a} and \ref{fig:Four b}], the absorption reduction at the line center is very small when $\Lambda_{12} = 0$.  Furthermore, inclusion of moderate incoherent pumping, for this case,  does not give rise to significant change in the EIT feature.   Whereas at higher values of $\Lambda_{12}$, there is reduction in overall absorption over a wider range of frequencies due to saturation of level populations as seen previously for the case of unequal decay rates.

       Figures \ref{fig:Five a} and \ref{fig:Five b} reveal that the intensity - intensity correlation function remains largely non classical for the range of pump Rabi field strengths we have studied, for the case of equal spontaneous decay rates.  The correlation function is seen to exhibit sub-poissonian characteristics mostly except for the largest value of control Rabi field of  $\Omega_{c} = \gamma_{21}$ considered here,  in which case there is a small window of cross-over into the classical regime from $ \tau $= 2.5 to 4  (where it is slightly greater than 1).   For times beyond this window of classicality, the correlation function is seen to approach the  steady state behaviour of uncorrelated random  field ($G_{22} (\infty)$  = 1).

\section{CONCLUSION}

  In conclusion, we have investigated the role of incoherent pumping on  photon statistics of the probe transition in a three level cascade  EIT configuration.  Both equal and  unequal spontaneous emission decay rates from upper levels, the latter of which corresponds to the actual experimental conditions in alkali systems were considered.  Our study demonstrates that although the incoherent pumping into the upper level of the probe transition doesn't  markedly alter the probe absorption (EIT ) characteristics,  it is seen to modify the second order correlation function of the probe field significantly.

  In the absence of any incoherent pumping ($\Lambda_{12}=0$), the function   $G_{22}(\tau)$   exhibits classical behaviour and it reaches the steady state of uncorrelated thermal light  within a few spontaneous emission cycles of the probe transition.   However, introduction of a small amount of incoherent pumping to the intermediate level causes the correlation function to attain  non classical character.
In addition, one notices that the influence of the incoherent pumping  is more during initial times and also that the correlation function reaches steady state earlier as is expected because of the nature of this additional interaction which decoheres the system.

 It is to be noted that,  at higher values of incoherent pumping, the absorption reduction occurs due to saturation of the populations of the levels participating in the transition as has been ascertained from the steady state level populations.

 The present study clearly demonstrates that a comprehensive understanding of the incoherent pumping can be attained from a study of the intensity-intensity correlation function rather than looking at the absorption profile alone.   Such a study may not only lead to a better understanding of the role of different processes  but   also provide  us   with a means of controlling the probe emission characteristics via manipulation of the system and   field parameters.

\section{ACKNOWLEDGEMENTS}
Shaik Ahmed acknowledges the University Grants Commission, India  and Preethi N. Wasnik acknowledges  RGNJRF(SC), University Grants Commission, India for financial support.
 

\begin{thebibliography}{99}
 \bibitem{r0}{H. Paul, Rev. Mod. Phys. 54, 1061-1102(1982); R. Loudon, 'The Quantum Theory  of Light', 2nd Ed. Oxford Univ. Press (1983).}
 \bibitem{r1}{H. Huang, S.-Y. Zhu, and M.S. Zubairy, M.O. Scully Phys. Rev. A {\textbf  {53}}, 1834 (1996).}
\bibitem{r2}{S.Swain, P. Zhou and  Z. Ficek, Phys. Rev.  A \textbf{61}, 043410(2000).}
\bibitem{r3}{Z. Ficek and S. Swain, Phys. Rev.  A {\textbf{ 69}}, 023401(2004).}
\bibitem{r4}{Sumanta Das and G.S. Agarwal,  Phys. Rev. A \textbf{77}, 033850(2008).}
\bibitem{r5}{G.S. Agarwal, J. von Zanthier, C. Skornia, and H. Walther,  Phys. Rev. A {\textbf { 65}}, 053826 (2002).}
\bibitem{r10}{C. Yao, W. Fei, S. Wen-Xing and X. Ming, Chin. Phys. Lett. \textbf{29}, 114209 (2012).}
\bibitem{r6}{ J.P. Marangos, J. Mod. Opt. \textbf{45}, 471, (1998); M. Fleischhauer, A. Imamoglu and J.P. Marangos, Rev. Mod. Phys. \textbf{77}, 633 (2005).}
\bibitem{r7}{Y. Zhu, Phys. Rev. A \textbf {45}, R6149 (1992);  M. Fleischhauer, C.H. Keitel, M.O. Scully and C. Su, Opt. Commun. \textbf {87} 109 (1992); J. Mompart and R. Corbalan, J.Opt.B: Quant. Semiclass. Opt. {\textbf 2} R 7 (2000).}
\bibitem{r8}{Suneel Singh and  M. Anil Kumar  Phys. Rev. A\textbf { 79}, 063821 (2009).}
\end{thebibliography}
\end{document}